\documentclass[11pt]{article}

\usepackage[margin=1in]{geometry}
\usepackage{setspace}
\usepackage[T1]{fontenc}
\usepackage[utf8]{inputenc}
\usepackage{lmodern}
\usepackage{textcomp}
\usepackage{microtype}

\usepackage{amsmath,amssymb,amsthm}
\usepackage{bm}
\usepackage{array}
\usepackage{booktabs}
\usepackage{calc}
\usepackage{caption}
\usepackage{subcaption}
\usepackage{enumitem}
\usepackage{float}
\usepackage{graphicx}
\usepackage{longtable}
\usepackage{multirow}
\usepackage{multicol}
\usepackage{threeparttable}
\usepackage[boxed]{algorithm2e}
\usepackage{xcolor}
\usepackage{comment}

\makeatletter
\def\maxwidth{\ifdim\Gin@nat@width>\linewidth\linewidth\else\Gin@nat@width\fi}
\def\maxheight{\ifdim\Gin@nat@height>\textheight\textheight\else\Gin@nat@height\fi}
\makeatother
\setkeys{Gin}{width=\maxwidth,height=\maxheight,keepaspectratio}

\newcommand{\spacingset}[1]{\renewcommand{\baselinestretch}{#1}\selectfont}

\makeatletter
\@ifundefined{c@theorem}{\newtheorem{theorem}{Theorem}}{}
\@ifundefined{c@lemma}{\newtheorem{lemma}{Lemma}}{}
\@ifundefined{c@assumption}{\newtheorem{assumption}{Assumption}}{}
\@ifundefined{c@remark}{\theoremstyle{remark}\newtheorem{remark}{Remark}}{}
\makeatother

\IfFileExists{marquis_general.sty}{\usepackage{marquis_general}}{}
\usepackage{notations}

\usepackage{xr}
\newcommand{\csection}[1]
    {\begin{center}
        \stepcounter{section}
        {\bf\large\arabic{section}. #1}
    \end{center}
}

\newcommand{\csubsection}[1]{
\begin{center}
\stepcounter{subsection}
{\it\arabic{section}.\arabic{subsection}. #1}
\end{center}
}

\renewcommand{\baselinestretch}{1.75}

\usepackage[round,authoryear]{natbib}
\usepackage{xurl}
\usepackage[colorlinks=true,linkcolor=blue,citecolor=blue,urlcolor=blue]{hyperref}
\usepackage{bookmark}

\hypersetup{
  pdftitle={An Instrumental Variable Approach to Account for Informative Treatment Switching in Real-world Evidence},
  pdfauthor={Yang Liu; Andrew Ying; Zongqi Xia; Jue Hou},
  pdfkeywords={causal inference; doubly robust; machine learning; survival analysis; electronic health records},
  pdfcreator={LaTeX}
}

\title{An Instrumental Variable Approach to Account for Informative Treatment Switching in Real-world Evidence}

\author{
Yang Liu\\
\small Division of Biostatistics and Health Data Science, School of Public Health, University of Minnesota\\
\small \texttt{liu02528@umn.edu}
\and
Andrew Ying\\
\small Department of Statistics and Data Science, The Wharton School, University of Pennsylvania
\and
Zongqi Xia\\
\small Department of Neurology, School of Medicine, University of Pittsburgh
\and
Jue Hou\\
\small Division of Biostatistics and Health Data Science, School of Public Health, University of Minnesota
}

\date{}

\begin{document}
\begin{center}
  {\bf\Large An Instrumental Variable Approach to Account for Informative Treatment Switching in Real-world Evidence}

  Yang Liu$^{1}$, Andrew Ying$^{2}$, Zongqi Xia$^{3}$, Jue Hou$^{1}$

{\it\small
$^1$ Division of Biostatistics and Health Data Science, School of Public Health, University of Minnesota;\\
$^2$ Google Inc.;\\
$^3$ Department of Neurology, School of Medicine, University of Pittsburgh.
}

\end{center}

\begin{singlespace}
\begin{abstract}
Reproducible and generalizable assessment of  treatment decisions requires principled handling of subsequent treatment switching that may inform expected outcomes and shift across cohorts and over time.
To effectively account for informative treatment switching, we propose an instrumental variable approach that characterizes the poorly documented expected outcomes at switching as unmeasured confounding.
After establishing the baseline treatment as a viable instrumental variable, we constructed an estimating equation based on the association between the centered instrumental variable and a martingale style residual process that identifies the treatment effect under structural cumulative survival model.
Our proposed method is doubly robust, i.e., valid whenever either of baseline propensity model or no-switching outcome model is consistently estimated.
A co-training of treatment effect parameter and survival outcome regression model eliminated the requirement of observing a no-switching subset under semi-parametric additive hazards models.
We further developed an baseline-survival-corrected cross-fitting approach to incorporate general machine learning models for estimating nuisance models.  Numerical results demonstrated the validity of our method in various settings when a basket of benchmark solutions produced biased or contradictory results. We applied our method to comparison of high-efficacy vs standard efficacy disease modifying treatments as the second line therapy of multiple sclerosis.
\end{abstract}


\end{singlespace}

\csection{Introduction}\label{sec-intro}

While clinical trials remain the gold standard for clinical evidence, EHR-derived real-world data has emerged as an increasingly important complement for addressing gaps in clinical knowledge, offering broader patient representation and long-term outcome that are not always feasible in trials due to their high costs, limited follow-up, and stringent eligibility criteria \citep{Franklin2021-pl, Franklin2020-nv}.
Proper adjustment for confounding is essential in observational studies to avoid confounding bias, yet remains challenging due to difficulties in identifying, measuring, and modeling the underlying factors \citep{Beaulieu-Jones2020-gr}.
In studies of continuous treatments, e.g. medications for chronic diseases administered in revolving cycles, one source of confounding often overlooked is treatment discontinuation and switching during the follow-up \citep{Hernan.Robins2006}. Reasons for treatment switching include lack of treatment efficacy \citep{Clephas.Malgie.ea2023}, adverse side effects \citep{Roussidis.Kalkavoura.ea2013}, emerging clinical evidence favoring newer options \citep{liang_temporal_2022} or accessibility of alternative therapies \citep{DegliEsposti.Favalli.ea2016}. Some of these factors are also associated with the observed and projected outcomes of interest, thus inducing confounding. Unfortunately, systematic documentation of these factors is often absent in EHRs, causing informative treatment switching with inadequately measured confounding \citep{Boyd.Paolino.ea2024}. In this study, we aim to develop estimation method for  treatment effects on time-to-event outcomes that specifically addresses the informative treatment switching.

Non-compliance, defined as deviation from the randomized treatment assignment, including treatment switching, has been extensively studied in trials via two common approaches, intention-to-treat (ITT) and per-protocol (PP) analyses \citep{Tripepi.Chesnaye.ea2020}.
ITT targets the effectiveness of treatment assignment regardless of adherence, but dilutes the actual treatment effect when switching is prevalent \citep{Hernan.Hernandez-Diaz2012}. PP targets treatment efficacy by restricting to compliant participants, yet may be biased if switching is informative for poor prognosis \citep{Murray.Caniglia.ea2021}.
Both approaches extend to observational studies \citep{Hou.Kim.ea2021}, where additional challenges arise.
ITT estimates are further limited in generalizability due to dynamic shifts in switching patterns across time and institutes.
Na\"{i}ve PP risks sampling bias by excluding switchers, who are often surrogates for adverse outcomes \citep{TTD_JCO18}.
Addressing these issues requires adequate measurement and adjustment for both pre-treatment and post-initiation confounders \citep{hernan_per-protocol_2017}, which can be challenging in complex real data analysis.

Existing literatures have addressed the treatment switching in causal estimation by modeling counterfactual survival times or treatment switching times.
Structural modeling approaches have been proposed to adjust for treatment switching, when time-dependent confounding is assumed to be fully measured \citep{Vansteelandt.Joffe2014a}.
The rank-preserving structural failure time model \citep{Robins.Tsiatis1991}, originally proposed to handle non-random treatment switching in randomized trials, represents each patient's counterfactual no-treatment failure time as a monotone transformation of observed time according to treatment history. However, its extension to observational data requires additional assumptions to handle confounding.
Alternatively, \cite{Robins.Finkelstein2000} treats switching as informative censoring and reweights observations to emulate full compliance, but requires all time-varying confounders associated with switching to be fully measured.
When the dynamic factors for treatment switching are inadequately documented, adoption of existing methods may produce a biased causal estimation due to violation on their assumptions.

Instrumental variable (IV) methods offer a well-established approach to recover treatment effects in presence of unmeasured confounding, and have been applied to account for informative treatment switching with potential unmeasured time-varying confounding in clinical trials
\citep{Papazoglou.Waddingham.ea2025}. The random assignment in a clinical trial is a plausible candidate for IV that satisfies its three core assumptions: (1) relevance -- it is associated with the treatment received by guiding treatment at baseline, (2) independence -- under double-blind conditions, it is independent of unmeasured prognostic factors, and (3) exclusion restriction -- it influences the outcome only through the actual treatment received as a decision on paper.
Several IV estimators have been developed for time-to-event outcomes in trials, including the contamination-adjusted ITT \citep{Sussman.Hayward2010}, structural cumulative survival model \citep{Martinussen.Vansteelandt.ea2017, ying_structural_2023}, and structural nested cumulative survival time model \citep{Seaman.Dukes.ea}. Extending these methods to observational studies requires identifying a suitable IV without randomized assignment.
For a point treatment under no unmeasured confounding, the observed treatment is conditionally randomized according to the true propensity score \citep{Rosenbaum.Rubin1983}. We propose to use the initial treatment to construct an IV through modeling its propensity score. Characterizing the confounding at baseline for the initial treatment is much more feasible than measuring time-varying confounding over the full follow-up.

Augmenting the propensity score approach by models of the potential outcomes can improve its efficiency when both models are consistently estimated or robustness when either model is misspecified.
Doubly robust estimators  \citep{Robins.Rotnitzky.ea1994,VanDerLaan.Rubin2006} combine outcome regression and weighting to achieve robustness against model misspecification. Recent extensions incorporate machine learning for flexible nuisance estimation \citep{Chernozhukov.Chetverikov.ea2018b,Hou.Bradic.ea2021}.
We aim to maintain the similar properties when relaxing from their non unmeasured confounding assumption with our IV method. A key challenge here is to recover the measured effects in outcome models while accounting for unmeasured time-varying confounding.

In this paper, we propose \textbf{DRIVE} (\textbf{D}oubly \textbf{R}obust \textbf{I}nstrumental \textbf{V}ariable \textbf{E}stimation), a method for estimating treatment effects on time-to-event outcomes in observational studies with informative treatment switching and unmeasured time-varying confounding. We established the role of initial treatment decision as an IV in suitable observational studies: it influences subsequent exposure trajectories but, under standard exclusion restrictions, affects the event time only through the administered therapy. DRIVE augments an propensity score based IV estimating equation with a marginal outcome-regression term, achieving unbiased estimation against misspecification of either model. We estimate the marginal outcome regression in two ways, (1) a one-shot semi-parametric  approach, and (2) a baseline-corrected machine-learning procedure guarding against the potential missing covariate bias in outcome models due to unmeasured time-varying confounding.
We list our key contributions herein:
\begin{enumerate}
\item We formulate the informative switching as an unmeasured confounding issue between time-varying treatment and time-to-event outcomes.
\item We characterize the initial treatment as the viable source of an  IV for time-varying treatment in presence of unmeasured post-initiation confounding.
\item We proposed novel methods for recovering measured hazards of potential outcomes correcting for unmeasured confounding, which achieve double robustness under semi-parametric models and asymptotic normality with  machine learning estimations.
\end{enumerate}

The remainder of the paper is organized as follows. Section \ref{aux-sec-notation} introduces the causal framework, IV, and doubly robust score function. Section \ref{aux-sec-method} presents both the semi-parametric and machine-learning-based estimators. Section \ref{aux-sc:theorem} establishes the asymptotic properties of the proposed method. Section \ref{aux-sec-sim} presents extensive simulation experiments. Section \ref{aux-sec-data} applies DRIVE to the comparative effectiveness analysis of high-efficacy versus standard-efficacy disease modifying treatments (DMTs) for multiple sclerosis (MS).

\csection{Setting and identification}\label{sec-notation}

In this section, we introduce the causal framework, identification assumptions, and IV-based doubly robust score for the treatment effect parameter, with notation defined in Section \ref{aux-sec-notation1}, causal setup in Section \ref{aux-sec-causal}, and identification in Section \ref{aux-sec-IV}.

\csubsection{Notation for observed data}\label{sec-notation1}

We consider $n$ independent observations indexed by $i = 1, \ldots, n$ with pre-treatment covariates $\Covs_i \in \Real^p$. Under continuous treatment switching, the treatment is a time-varying process $\Treat_i(\TimeGrid) \in \{0, 1\}$ for $\TimeGrid \in [0,\TimeMax]$ up to the maximal follow-up time $\TimeMax$. We denote the observed treatment history before time $\TimeGrid$ as $\TreatHist_i(\TimeGrid) = \{\Treat_i(s): 0\leq s\leq \TimeGrid\}$ and cumulative treatment time as $\cumTrt_i(\TimeGrid) = \int_0^{\TimeGrid-} \Treat_i(s)\,ds$. We denote $\TimeSurv_i$ as the event time of interest.
Under standard right censoring at $\TimeCens_i$, we observe  the event indicator as $\IndEvent_i = \Indicator(\TimeSurv_i < \TimeCens_i)$ and $\TimeObs_i = \TimeSurv_i \wedge \TimeCens_i$ where we adopt the notation $a\wedge b = \min\{a, b\}$ as observation time. Thus, our observed data consist of $\left\{\TimeObs_i, \IndEvent_i, \TreatHist_i(\TimeObs_i), \Covs_i\right\}$ for $i = 1, ..., n$. In the following, we abbreviate $\TreatHist_i(\TimeObs_i)$ simply as $\TreatHist_i$.
We adopt the counting process notations \citep{Andersen.Gill1982a} for event $\ProcCount_i(\TimeGrid) = \IndEvent_i \Indicator(\TimeObs_i < \TimeGrid)$ and at-risk $\ProcRisk_i(\TimeGrid) = \Indicator(\TimeObs_i \ge \TimeGrid)$.

\csubsection{Potential outcomes and IV}\label{sec-causal}

We approximate the continuous time domain $\TimeGrid \in [0,\TimeMax]$ by the discrete partition $0=\TimeGrid_0<\TimeGrid_1<\dots<\TimeGrid_M = \TimeMax$, the time resolution of potential treatment changes at discrete clinical visits.
A treatment process $\trt(\TimeGrid)$ defined on the partition is the vector $\vtrt = (\trt_0,\dots,\trt_M)$ with $\trt_m = \trt(\TimeGrid_m)$. The potential event time $\TimePot$ is defined over full sequence of treatment $\vtrt$. Similarly, we have the potential counting process $\PotCount{\TimeGrid} = \IndEvent_i \Indicator\left\{ \TimePot \wedge \TimeCens_i \le  \TimeGrid \right\}$ and at-risk process $\PotRisk{\TimeGrid} = \left\{ \TimePot \wedge \TimeCens_i \ge  \TimeGrid \right\}$. We denote the vector of observed treatment history up to $\TimeGrid_m$ as $\TreatHist_i(\TimeGrid_{m}) = \left(\Treat(\TimeGrid_0),\dots,\Treat(\TimeGrid_m)\right)^\top$, constant $m$-dimensional vectors of ones $\trted{m}$ and zeros $\ctrl{m}$ to construct potential event time with truncated treatment sequence, $\Tblip[m] = \TimePot[\TreatHist_i(\TimeGrid_m), \ctrl{M-m}]$.
The treatment effect $\EffParam$ is characterized by contrasting $\Tblip[m]$ and $\Tblip[m-1]$ different by a blip at $\TimeGrid_m$ through a structural cumulative survival model (SCSM) for $m = 1,\dots,M -1$,
\begin{equation}
\begin{aligned}
\label{eq:SCSM}
\frac{\Prob\left[\Tblip[m]>\TimeGrid \mid \TreatHist_i\left(\TimeGrid_m\right), \Covs_i, \TimeSurv_i \geq \TimeGrid_m\right]}{\Prob\left[\Tblip[m-1] >\TimeGrid \mid \TreatHist_i\left(\TimeGrid_m\right), \Covs_i, \TimeSurv_i \geq \TimeGrid_m\right]}
=\exp \left\{-\int_{\TimeGrid_m}^{\TimeGrid \wedge \TimeGrid_{m+1}} \EffParam \Treat_i\left(\TimeGrid_m\right) ds\right\}.
\end{aligned}
\end{equation}
The treatment effect parameter $\EffParam$ in SCSM~\eqref{aux-eq:SCSM} is the hazard difference in $[\TimeGrid_m, \TimeGrid_{m+1}]$ among the treated in the window, i.e. $\Treat(\TimeGrid_m) = 1$, between two potential outcomes with truncated treatment up to $\TimeGrid_m$ or $\TimeGrid_{m+1}$. In the current study, we focus on the constant $\EffParam$ without heterogeneity over time $\TimeGrid$ or baseline covariates $\Covs_i$ \citep{Hou.Bradic.ea2021}.  The model~\eqref{aux-eq:SCSM} also implicitly leaves the treatment effect unspecified for the control in the window, i.e. $\Treat(\TimeGrid_m) = 0$, and assumes same hazard function for the two potential outcomes beyond $\TimeGrid_{m+1}$.

In randomized clinical trials with non-compliance, the initial treatment assignment, or randomization, has been used for IV under non-compliance \citep{cuzick_adjusting_1997} or non-adherence \citep{ying_structural_2023}. In observational studies, if baseline covariates adequately measure the confounding, the initial treatment assignment behaves like conditional randomization \citep{Rosenbaum.Rubin1983}. Therefore, we propose to investigate the role of the initial treatment as an IV, $\IV_i = \Treat_i(0)$, to bypass the post-initiation unmeasured confounding that induces the informative treatment switching.
To conceptually justify the proposed IV, we inspect the three standard conditions. For the \emph{relevance} condition, the initial treatment choice naturally affects the subsequent time-varying treatment. For the \emph{independence} condition, whenever pre-treatment covariates fully explains the projected prognosis considered at baseline, the initial treatment decision should be conditionally independent of post-initiation prognosis inducing informative switching. For the \emph{exclusion restriction} condition, there is usually no biological or clinical mechanism through which the initial treatment order, beyond the treatment actually received over time, would directly influence patient outcomes.
We provide the formal assumptions and the identifiability of treatment effect $\EffParam$ through an IV approach in the next section.

\csubsection{Assumptions and identification through DRIVE score}\label{sec-IV}

To identify the treatment effect $\EffParam$ under SCSM~\eqref{aux-eq:SCSM}, we impose the following assumptions that specifies the typical consistency, non-informative censoring and the IV assumptions (relevance, independence and exclusion restriction) under our setting.
\begin{assumption}[Causal setting]\label{ap:causal} We make the following assumptions throughout the paper with an absolute constant $\varepsilon>0$.
  \begin{enumerate}[label=(\roman*), ref=\ref{aux-ap:causal}(\roman*)]
      \item \label{ap:consistency} \textbf{Cumulative consistency}: For any $m\in\{0,\ldots,M-1\}$,
      $
      \Tblip[m]\wedge \TimeGrid_{m+1}
      =
      \TimeSurv_i\wedge \TimeGrid_{m+1}
      $.

      \item \label{ap:censor} \textbf{Non-informative censoring}:
        $\TimeCens_i \Indep
        \left\{
        \Tblip[m],\Treat_i(\TimeGrid_m): m=0,\dots,M
        \right\}
        \mid \Covs_i$.

   \item \label{ap:positivity} \textbf{Adequate at-risk}. \(
\Exp\{\ProcRisk_i(\TimeMax)\}\ge \varepsilon.
\)
    \item \label{ap:IVr} \textbf{IV Relevance}: The IV $\IV_i$ is correlated with the treatment history after adjusting for baseline covariates (See Section~\ref{aux-sec:supp_relevance} of Supplementary Materials).
    \item  \label{ap:IVi}  \textbf{Independence}: The IV is conditionally independent of no-switching control event time,
		$\IV_i \Indep \TimePot[\ctrl{M}] \mid \Covs_i$.
    \item  \label{ap:IVer}  \textbf{Exclusion restriction}: IV has no direct effect on the outcome except through the exposure,
    $\TimePot[0,\trt_1,\dots,\trt_M] = \TimePot[1,\trt_1,\dots,\trt_M]$.
  \end{enumerate}
\end{assumption}

Assumption~\ref{aux-ap:consistency} enhances the standard consistency by the compatibility over time --- the event time up to $\TimeGrid_{m+1}$ should be the same if the hypothetical treatment truncation only happens after $\TimeGrid_{m+1}$. Under Assumption~\ref{aux-ap:consistency}, we can characterize the informative treatment switching under SCSM~\eqref{aux-eq:SCSM} as the dependence between potential outcomes of the control and observed treatment (see Section~\ref{aux-sec:supp_consistency} of Supplementary Materials).
Assumption~\ref{aux-ap:censor} imposes conditional non-informative censoring as a missing-at-random for causal identification \citep{Ding.Fan2018}.
In the EHR applications, censoring is the termination of current care, which is typically caused by exogenous reasons unrelated to post-initiation clinical factors. Assumption~\ref{aux-ap:positivity} is a standard assumption in survival analysis that ensures the stability of estimating various baseline cumulative hazards functions \citep{Martinussen.Vansteelandt.ea2017,LinYing94}, which can always be satisfied by adjusting last follow-up time $\TimeMax$.

Assumptions~\ref{aux-ap:IVr}--\ref{aux-ap:IVer} formalize the three standard conditions for IV discussed in Section~\ref{aux-sec-causal}.
Assumption~\ref{aux-ap:IVr} is the relaxed from the point-wise relevance in \citet{ying_structural_2023}, enabled by the time-invariant treatment effect $\EffParam$ under SCSM~\eqref{aux-eq:SCSM}. Unlike the guaranteed relevance in randomized studies whose control has no access to treatment, initial treatment in real-world care may have a diminishing point-wise relevance over time.
The Assumption~\ref{aux-ap:IVi} is the standard ignorability condition of initial treatment for treatment effect of the treated problem \citep{Robins.Tsiatis1991}, as the SCSM~\eqref{aux-eq:SCSM} implicitly focuses on the observed treated subjects.
The Assumption~\ref{aux-ap:IVer} eliminates any direct effect of initial treatment on the potential outcomes.

To identify the treatment effect, we leverage two nuisance models through a doubly robust score.
Besides the initial propensity score, $\PScore(\Covs_i) = \Prob(\IV_i=1\mid \Covs_i)$, we leverage another working outcome models.
Define the conditional hazard and cumulative hazard functions of no-switching control outcome $\TimePot[\ctrl{M}]$,
\begin{equation}\label{def:control-haz}
    \hazctrl(\TimeGrid;\Covs_i) = \lim_{\Delta \to 0+} \Prob\left(\TimePot[\ctrl{M}]< \TimeGrid + \Delta \mid \Covs_i,\TimePot[\ctrl{M}]\ge \TimeGrid\right)/\Delta, \;
    \HazMeas(\TimeGrid;\Covs_i) = \int_0^{\TimeGrid} \hazctrl(s;\Covs_i)ds.
\end{equation}

For each treatment period $\tperiod \in [\TimeGrid_m, \TimeGrid_{m+1})$, we propose the working model with additive unmeasured confounding $\ConfUn_i(\TimeGrid)$,
\begin{align}
\lambda\left(\tperiod \mid\TreatHist_i(\tperiod), \Covs_i, \ConfUn_i(\tperiod)\right)
= & \lim_{\Delta \to 0+} \Prob\left(\TimeSurv_i< \tperiod + \Delta \mid  \TreatHist_i(\tperiod),\Covs_i,\ConfUn_i(\tperiod), \TimeSurv_i\ge \tperiod\right)/\Delta \notag \\
= & \EffParam \cdot\Treat_i(\TimeGrid_m) + \hazctrl\left(\tperiod;\Covs_i\right) +\ConfUn_i(\tperiod).  \label{eq:SurvModel}
\end{align}
Through the working model~\eqref{aux-eq:SurvModel}, we characterize the measured confounding as
$\hazctrl\left(\TimeGrid;\Covs_i\right)$
and the unmeasured confounding inducing the informative treatment switching as $\ConfUn_i(\TimeGrid)$.

Using the centered IV, $\IV_i - \PScore(\Covs_i)$, and the martingale style residual for survival outcome, $$
\ResidMart_i\{\TimeGrid;\EffParam, \HazCum\} = \ProcCount_i(\TimeGrid) - \int_0^{\TimeGrid}\ProcRisk_i(s)\left\{\EffParam \Treat_i(s)ds+d\HazCum(s;\Covs_i)\right\},$$
we construct the DRIVE score
\begin{equation}
		\label{eq:ScoreFunction}
		\Psi_i\left\{\EffParam;\PScore, \HazCum\right\} = \int_0^{\TimeMax} \exp\left\{\EffParam\cumTrt_i(\TimeGrid-)\right\}
		\left\{\IV_i - \PScore(\Covs_i)\right\} d\ResidMart_i\{\TimeGrid;\EffParam, \HazCum\}.
\end{equation}
\begin{remark}\label{remark:exp-score}
The exponential weighting term $\exp\left\{\EffParam \cumTrt_i(\TimeGrid-) \right\}$ and the linear treatment component $\EffParam \Treat_i(\TimeGrid)$ negates the impact of treatment on both the at-risk process $\ProcRisk_i(\TimeGrid)$ and the counting process $\ProcCount_i(\TimeGrid)$ under Model~\eqref{aux-eq:SCSM} \citep{Hou.Bradic.ea2021,dukes2024doubly} so that their conditional expectations would match those of their counterparts for non-switching control, $\PotRisk[i,\ctrl{M}]{\TimeGrid}$ and $\PotCount[i,\ctrl{M}]{\TimeGrid}$ as defined in Section \ref{aux-sec-causal} (see Section~\ref{aux-sec:supp_consistency} and  Lemma~\ref{aux-lem:N-reduction} in Section~\ref{aux-sec-aux} in Supplementary Materials).
\end{remark}

Let $\EffTrue$, $\PStrue$ and $\Haztrue$ be the true treatment effect and nuisance models, respectively.
We established the identifiability of treatment effect under SCSM~\eqref{aux-eq:SCSM} via the DRIVE score~\eqref{aux-eq:ScoreFunction}.
\begin{lemma}\label{lm:identonestep}
Under Assumption~\ref{aux-ap:causal}, the DRIVE score~\eqref{aux-eq:ScoreFunction} identifies the true effect $\EffTrue$ when at least one of following is satisfied: (1) $\PScore = \PStrue$, or (2) $\HazCum = \Haztrue$,
$$
\Exp\left[\Psi_i\left\{\EffTrue;\PScore,\HazCum \right\}\right]
  = \Exp\left(\int_0^{\TimeMax} \{\IV_i-\PScore(\Covs_i)\}\,[d\PotCount[i,\ctrl{M}]{\TimeGrid}-\PotRisk[i,\ctrl{M}]{\TimeGrid}d\HazCum(\TimeGrid;\Covs_i)]\right)=0.
$$
\end{lemma}
 The proof is provided in Section \ref{aux-sec:supp_lemma1} in Supplementary Material. We first formalized Remark~\ref{aux-remark:exp-score} that the conditional expectation of the martingale style residual with exponential weighting equals that of the potential residuals under no-switching control.
 Under Assumption~\ref{aux-ap:IVi} (IV independence) the two residuals for initial propensity score and no-switching control counting process are uncorrelated. Whenever either residual term has mean zero, their product has expectation zero.

Doubly robust estimation methods often possess the Neyman orthogonality, which enables the statistical inference of treatment effect using potentially non-regular estimators for nuisance models \citep{Chernozhukov.Chetverikov.ea2018b,Hou.Bradic.ea2021}.
We show that DRIVE  score~\eqref{aux-eq:ScoreFunction} also satisfy the Neyman orthogonality in the following sense.
\begin{lemma}[Neyman orthogonality]\label{lm:orthogonality}
Under Assumptions~\ref{aux-ap:causal}, the DRIVE score
  \(\Psi_i(\EffParam;\PScore,\HazCum)\)
  defined in \eqref{aux-eq:ScoreFunction} satisfies
  \[
  \left.\frac{\partial}{\partial r}
  \Exp\big[\Psi_i\{\EffTrue;
  \PStrue+r(\PScore-\PStrue),
  \Haztrue+r(\HazCum-\Haztrue)\}\big]\right|_{r=0}=0.
  \]
\end{lemma}
We proved the lemma using a similar strategy in the proof of Lemma~\ref{aux-lm:identonestep}, laid out in Remark~\ref{aux-remark:exp-score} (see
See Section~\ref{aux-proof:lem-neyman} in Supplementary Materials). Like in existing literatures, the property established in Lemma \ref{aux-lm:orthogonality} is only applicable when all nuisance parameters approach their true values \citep{Chernozhukov.Chetverikov.ea2018b,Hou.Bradic.ea2021}. Following the Neyman orthogonality, the estimation equation constructed according to the score $\Psi$ would be insensitive to estimation errors of nuisance models  $(\PScore,\HazCum)$. Conclusion of Lemma \ref{aux-lm:orthogonality} is the foundation for designing the DRIVE estimator with flexible machine learning and deriving its statistical inference.

\csection{Estimation and inference methods}\label{sec-method}

Lemma~\ref{aux-lm:identonestep} motivates our DRIVE estimation of $\EffParam$ by solving the empirical estimating equation
\begin{equation*}
\frac{1}{n}\sum_{i=1}^n \Psi_i\{\EffParam; \EstPScore, \EstHaz\} = 0,
\end{equation*}
with estimated nuisance models $\EstPScore$ and $\EstHaz$ for  $\PStrue$ and  $\Haztrue$, respectively.
In presence of unmeasured confounding,
estimation of the measured confounding effect $\Haztrue(\cdot)$ is non-trivial and requires additional assumptions for the working model~\eqref{aux-eq:SurvModel} (see Section~\ref{aux-sec:supp_ml_bias} in Supplementary Materials).
We provide two strategies: (i) under a semi-parametric specification via a system of joint estimating equations (Section \ref{aux-sec-parametric}); and (ii) through a co-training scheme that integrates machine-learning estimators with a baseline hazard correction (Section \ref{aux-sec-ml}).

\csubsection{Joint estimation with semi-parametric specification}\label{sec-parametric}

Suppose we impose parametric and semi-parametric working models for the propensity score and the hazard model, respectively,
\begin{equation}
    \PScore(\Covs_i)=\expit\left(\ParamGamma^{\top}\Covs_i\right), \,
    \expit(x)= 1/\left(1+e^{-x}\right),
\;
\HazMeas(\TimeGrid;\Covs_i)=\TimeGrid \cdot\ParamAlpha^{\top}\Covs_i+ \HazBase(\TimeGrid), \label{eq:linearspecification}
\end{equation}
with the corresponding martingale style residual
\begin{equation*}\ResidMart_i(\TimeGrid; \EffParam, \ParamAlpha, \HazBase) = \ProcCount_i(\TimeGrid) -\int_0^{\TimeGrid}\ProcRisk_i(s)\left[\left\{\EffParam \Treat_i(\TimeGrid)+\ParamAlpha^\top \Covs_i\right\} ds +d\HazBase(s)\right].\end{equation*}

We propose to construct an estimator that simultaneously recovers both the treatment effect $\EffParam$ and $\HazMeas$.
Applying the nuisance models~\eqref{aux-eq:linearspecification} to the DRIVE score~\eqref{aux-eq:ScoreFunction},we obtain the score
\begin{equation}
		\label{eq:ScoreFunctionjoint1}
		\Psi_i\left(\EffParam;\ParamAlpha , \ParamGamma,  \HazBase\right) = \int_0^{\TimeMax} \exp\left\{\EffParam\cumTrt_i(\TimeGrid-)\right\}
		\left\{\IV_i - \expit\left(\ParamGamma^\top \Covs_i\right)\right\} d\ResidMart_i(\TimeGrid;\EffParam, \ParamAlpha, \HazBase).
\end{equation}

To estimate the coefficients $\ParamAlpha$ of measured confounding effect $\HazBase$,
we construct another score function,
\begin{equation}
	\label{eq:ScoreFunctionjoint2}
	\Phi_i\left(  \ParamAlpha; \EffParam,\HazBase \right) = \int_0^{\TimeMax} \exp\left\{ \EffParam \cumTrt_i(\TimeGrid-) \right\} \Covs_i \, d\ResidMart_i(\TimeGrid; \EffParam, \ParamAlpha, \HazBase).
\end{equation}

For both \eqref{aux-eq:ScoreFunctionjoint2}~and~\eqref{aux-eq:ScoreFunctionjoint1},
we estimate the baseline function $\HazBase$ using the Breslow type estimator,
\begin{equation}
	\label{eq:SimpleBaseline}
	\EstBase(\TimeGrid; \EffParam, \ParamAlpha) = \int_0^{\TimeGrid} \frac{ \sum_i \exp\left\{ \EffParam \cumTrt_i(s-) \right\} \ProcRisk_i(s) \left[ d\ProcCount_i(s) - \left\{ \EffParam \Treat_i(s) + \ParamAlpha^\top \Covs_i \right\} dV_i(s) \right] }{ \sum_i \exp\left\{ \EffParam \cumTrt_i(s-) \right\} \ProcRisk_i(s) },
\end{equation}
where \( dV_i(\TimeGrid) = \sum_{m=1}^{M} (\TimeGrid_m - \TimeGrid_{m-1}) d\ProcCount_i(\TimeGrid) \) is a step function approximation used to simplify the computation of integrals.

After estimating $\hat\ParamGamma$ through the logistic regression, we combine \eqref{aux-eq:ScoreFunctionjoint1}, \eqref{aux-eq:ScoreFunctionjoint2} and \eqref{aux-eq:SimpleBaseline} to jointly estimate \(\EffParam,\ParamAlpha\) by solving the estimating equations
\begin{equation}\label{eq:jointscore}
  \frac{1}{n}\sum_{i=1}^n\Psi_i\left(\EffParam,\ParamAlpha;\hat\ParamGamma,\EstBase(\cdot;\EffParam,\ParamAlpha)\right)=0,
  \; \frac{1}{n}\sum_{i=1}^n\Phi_i\left(\EffParam,\ParamAlpha;\EstBase(\cdot;\EffParam,\ParamAlpha)\right)=0.
 \end{equation}

We denote the solution to \eqref{aux-eq:jointscore} as $(\EffEstJoint,\EstAlpha_{\mathrm{joint}})$.
Through standard inference method for estimating equations, we derive the closed form variance estimator $\hat{\mathcal V}_{\mathrm{joint}}$ (see Section~\ref{aux-sec:supp_jointVar} of Supplementary Materials).

\csubsection{ML estimation and bias correction}\label{sec-ml}

When covariates interact in a complex nonlinear manner to affect initial treatment or time-to-event outcome, the parametric and semi-parametric models \eqref{aux-eq:linearspecification} based on linear prediction of covariates can not fully characterize the confounding. We therefore consider flexible models to be estimated by machine-learning methods.
Under model \eqref{aux-eq:SurvModel}, the measured confounding effect $\HazMeas(\TimeGrid;\Covs_i)$ reflects the heterogeneity in outcomes along covariates $\Covs_i$ with controlled treatment.
Considering existing machine-learning methods do not support joint estimation like \eqref{aux-eq:jointscore}, we consider applying ML techniques on the no-switching control group, $\TreatHist_i = \ctrl{M}$, to estimate $\HazMeas(\TimeGrid;\Covs_i)$. \cite{Hou.Bradic.ea2021} proposed a estimation framework by combining the separate estimation of outcome survival model restricted to treated and control groups. Likewise, we estimate the survival function $\hat{S}_{\ctrl{M}}(\TimeGrid;\Covs_i)$ to obtain the cumulative hazards $\hat{\HazCum}^{\dag}(\TimeGrid;\Covs_i) = -\log\left\{\hat{S}_{\ctrl{M}}(\TimeGrid;\Covs_i)\right\}$.
Due to presence of unmeasured $\ConfUn_i(\TimeGrid)$ in working model~\eqref{aux-eq:SurvModel}, the ML estimator $\hat{\HazCum}^{\dag}$ is biased for the measured confounding effect $\HazMeas(\TimeGrid;\Covs_i)$.
Under suitable independence assumption between measured and unmeasured confounding effects, the bias is invariant along baseline covariates $\Covs_i$ (see Assumption~\ref{aux-ap:indepL-ml} and
Section~\ref{aux-sec:supp_ml_bias} of Supplementary Materials).
 To account for this bias that is otherwise difficult to estimate, we introduce a data-driven correction $\EstBase_r(t; \EffParam, \hat{\HazCum}^{\dag})$ to account for the bias with the proper treatment effect parameter $\EffParam$,
\begin{equation*}
	\EstBase_r(\TimeGrid; \EffParam, \hat{\HazCum}^{\dag}) = \int_0^{\TimeGrid}\frac{ \sum_i \exp\{\EffParam\cumTrt_i(\TimeGrid-)\} \left[ d\ProcCount_i(\TimeGrid) - \ProcRisk_i(\TimeGrid)\{\EffParam \Treat_i(\TimeGrid)  d\TimeGrid + d \hat{\HazCum}^{\dag}(\TimeGrid; \Covs_i)\} \right] }{ \sum_i \exp\{\EffParam\cumTrt_i(\TimeGrid-)\} \ProcRisk_i(\TimeGrid) },
\end{equation*}
and obtain the adjusted cumulative hazard estimator as:
\[
\EstHaz(\TimeGrid; \Covs_i) = \hat{\HazCum}^{\dag} (\TimeGrid; \Covs_i) + \EstBase_r(\TimeGrid; \hat{\EffParam},  \hat{\HazCum}^{\dag}).
\]
Considering highly flexible machine learning model can result in slower than parametric $n^{-1/2}$ estimation rate and potential overfitting bias, we adopt the cross-fitting co-training strategy to achieve $\sqrt{n}$ asymptotically normal estimation of $\EffParam$.
We randomly partition the data into $K$ folds, with $\mathcal{I}_1,\ldots,\mathcal{I}_K$ denoting the indices for in-fold data and  $\mathcal{I}_{-j}=\{1,\dots,n\}\setminus\mathcal{I}_j$ for out-of-fold data. For each fold $j$, we fit the nuisance functions on the training folds $\mathcal{I}_{-j}$, yielding \( \cfPScore \) and \( \hat \HazCum^{\dagger(j)}\) to obtain out-of-fold predictions for subjects $i \in \mathcal{I}_j$. Rather than estimating the in-of-fold specific correction, we pool these cross-fitted predictions across all $n$ subjects to calculate a single global correction component, $\EstBase_r(\TimeGrid;\EffParam,\hat \HazCum^{\dagger})$, which improves finite-sample stability by avoiding small risk sets at later time points. Finally, this global correction and the cross-fitted nuisance estimates are inserted into constructing the individual score contributions.
\begin{gather}
\Psi_i\left(\EffParam;\hat\PScore^{(j)},\hat\HazCum^{\dagger(j)},\EstBase_r\right)= \int_0^{\TimeMax} \exp\left\{\EffParam\cumTrt_i(\TimeGrid-)\right\}
		\left\{\IV_i - \hat\PScore^{(j)}(\Covs_i)\right\} d\ResidMart_i\left(\TimeGrid;\EffParam, \EstBase_r,\hat\HazCum^{\dagger(j)}\right), \notag \\
        \ResidMart_i\left(\TimeGrid;\EffParam, \EstBase_r,\hat\HazCum^{\dagger(j)}\right) =
        \ProcCount_i(\TimeGrid) -\int_0^{\TimeGrid}\ProcRisk_i(s)\left\{\EffParam \Treat_i(s) ds  + d\hat\HazCum^{\dagger(j)}(s; \Covs_i)+d\EstBase_r(s;\EffParam,\hat \HazCum^{\dagger(j)})\right\}. \label{def:ml-score}
\end{gather}
We construct the cross-fitted estimating equation by averaging \eqref{aux-def:ml-score} across all subjects to solve for $\EffEstML$,
\begin{equation*}\label{def:ee-ml}
    \frac{1}{n}\sum_{j=1}^K\sum_{i \in \mathcal{I}_j} \Psi_i\left(\EffParam;\hat\PScore^{(j)},\hat\HazCum^{\dagger(j)},\EstBase_r\right)  = 0.
\end{equation*}
We estimate the asymptotic variance of $\EffEstML$ by
\begin{gather*}
\hat{\mathcal{V}}_{\mathrm{ML}} = \frac{n\sum_{j=1}^K\sum_{i\in\mathcal I_j}\Psi_i^2(\EffEstML; \cfPScore(\cdot), \hat{\HazCum}^{(j)}(\TimeGrid; \Covs_i))}{\hat{\Sigma}_{\IV\Treat,\mathrm{ML}}^2}, \\
\begin{aligned}\hat{\Sigma}_{\IV\Treat,\mathrm{ML}} =&
 n^{-1}\sum_{j=1}^K\sum_{i\in\mathcal I_j}  \left[ \left\{\IV_i - \cfPScore(\Covs_i)\right\}\int_0^{\TimeMax}
\exp\left\{ \hat\EffParam_{\mathrm{ML}} \cumTrt_i(\TimeGrid-) \right\} \ProcRisk_i(\TimeGrid)\Treat_i(\TimeGrid)d\TimeGrid \right.\\
&+ \left.\int_0^{\TimeMax} \left\{\IV_i - \cfPScore(\Covs_i)\right\}\left\{\cumTrt_i(\TimeGrid-)\right\} \exp\left\{ \hat\EffParam_{\mathrm{ML}} \cumTrt_i(\TimeGrid-) \right\} \ProcRisk_i(\TimeGrid)d\ResidMart_i\left(\TimeGrid;\EffParam, \EstBase_r,\hat\HazCum^{\dagger(j)}\right)\right].
\end{aligned}
\end{gather*}

\csection{Theoretical Results}\label{sc:theorem}
We establish asymptotic normality and double robustness for both DRIVE estimators. The semi-parametric estimator (Section~\ref{aux-sec-theory-par}) is asymptotical normal if either nuisance models is correctly specified. The ML-based estimator (Section~\ref{aux-sec-theory-ml}) achieves asymptotic normality when the product of convergence rates of the two nuisance models exceeds $n^{-1/2}$ rate.

\csubsection{Semi-parametric Estimator}\label{sec-theory-par}
To distinguish the misspecified model from the true model, we define the limiting parameters $\LimGamma$, $\LimAlpha$, and $\LimBase$ for  $\EstGamma$, $\EstAlpha$, and $\EstBase(\cdot; \EffTrue, \TrueAlpha)$, respectively. These parameters coincide with the true values $\TrueGamma$, $\TrueAlpha$, or $\TrueBase$ when the corresponding model is properly specified. The specific definitions are given in Definition \ref{aux-def:limits} (Section \ref{aux-sec:supp_mainproof} in the Supplementary Materials).
For the DRIVE with semi-parametric models, we consider the following regularity conditions.
\begin{assumption}[Regularity for semi-parametric estimator]\label{ap:joint} For absolute constants  $K_{\Covs}<\infty$ and $\varepsilon > 0$,
\begin{enumerate}[label=(\roman*), ref=\ref{aux-ap:joint}(\roman*)]
    \item\label{ap:boundedcovariate} \textbf{Bounded covariates}.
\(
\Prob\!\left(\max_{1\le i\le n}\|\Covs_i\|_\infty < K_{\Covs}\right)=1.
\)
  \item \label{ap:regularity}
  \textbf{Invertibility}. The asymptotic Jacobian matrix is positive definite,
\[
\bv^\top \Exp
\begin{pmatrix}\left.
\partial_{\EffParam} \left\{\Psi_i(\EffParam,\ParamAlpha; \bar\ParamGamma,\EstBase(\cdot; \EffParam, \ParamAlpha))\right\}\right|_{(\EffParam,\ParamAlpha)=(\EffTrue,\bar{\ParamAlpha})}
\\
\left.\partial_{\EffParam} \left\{\Phi_i(\EffParam,\ParamAlpha;\EstBase(\cdot; \EffParam, \ParamAlpha))\right\}\right|_{(\EffParam,\ParamAlpha)=(\EffTrue,\bar{\ParamAlpha})}
\end{pmatrix}
\bv
\ge \varepsilon, \;
\forall \bv \in \Real^{p+1}, \, \|\bv\|_2 = 1.
\]
\item\label{ap:drjoint} \textbf{Nuisance mode specification}. At least one model in \eqref{aux-eq:linearspecification} is correctly specified
\begin{enumerate}[label=\alph*), ref = \ref{aux-ap:drjoint}-\alph*]
    \item \label{ap:drjoint-ps} \underline{Initial propensity score}.   $\PStrue(\Covs) = \expit(\TrueGamma^\top\Covs_i)$;
    \item \label{ap:drjoint-or} \underline{Hazard model}.
    $
\Haztrue(\TimeGrid;\Covs_i) = \TimeGrid \cdot \TrueAlpha^\top\Covs_i + \TrueBase(\TimeGrid)$, and $   \{\ConfUn_i(\TimeGrid): \TimeGrid\in[0,\TimeMax]\} \Indep \Covs_i \mid \IV_i.
    $
\end{enumerate}
\end{enumerate}

\end{assumption}

Assumption~\ref{aux-ap:boundedcovariate} is standard regularity condition in survival analysis \citep{Martinussen.Vansteelandt.ea2017,LinYing94}.
Assumption~\ref{aux-ap:regularity} is a local nonsingularity condition for the joint estimating equation. It ensures local uniqueness of the solution.
Assumption~\ref{aux-ap:drjoint-or} characterizes a sufficient independent condition between measured and unmeasured confounding effects under which the measured confounding effect can be jointly estimated with treatment effect.

We now present the asymptotic properties of our estimator $\EffEstJoint$ with parametric propensity score and semi-parametric outcome regression models solved from \eqref{aux-eq:jointscore}.

\begin{theorem}\label{thm:joint}
Under Assumptions~\ref{aux-ap:causal}~and~\ref{aux-ap:joint},
we have
\[
\sqrt{n}\,(\EffEstJoint - \EffTrue)
\;\;\rightsquigarrow\;\;
N\!\left(0,\, \mathcal{V}_{\mathrm{joint}}\right)
\text{ and }
\sqrt{n/\hat{\mathcal{V}}_{\mathrm{joint}}}\,(\EffEstJoint - \EffTrue)
\;\;\rightsquigarrow\;\;
N\!\left(0,\, 1\right).
\]
\end{theorem}
From Lemma \ref{aux-lm:identonestep}, we already know that the treatment effect $\EffTrue$ can be identified by the correct propensity score model and an arbitrary outcome model.
To prove Theorem \ref{aux-thm:joint}, the key is to show for the other situation that DRIVE can simultaneously estimate treatment effect $\EffTrue$ and outcome model parameter $\TrueAlpha$ without consistent propensity scores,
for which we leveraged the Assumption~\ref{aux-ap:drjoint-or}.  Combining with the conclusion of Lemma \ref{aux-lm:identonestep}, the joint estimating equations \eqref{aux-eq:jointscore} hence identify the $\EffTrue$ and $\TrueAlpha$ simultaneously.  Detailed proof is provided in Section \ref{aux-proof:thm_joint} in the Supplementary Material.


A distinctive feature of DRIVE's semi-parametric estimator is the joint estimation of the treatment effect $\EffTrue$ and the outcome-model coefficient $\TrueAlpha$ through  \eqref{aux-eq:jointscore}. In conventional doubly robust procedures, such as the augmented inverse probability weighting (AIPW) estimator \citep{Robins.Rotnitzky.ea1994} and its semi-parametric survival extensions \citep{Wang.Lee.ea2017, Dukes.Martinussen.ea2019}, the nuisance models are typically estimated in a separate first step and then plugged into the treatment-effect score. This sequential strategy requires that outcome model can be independently estimated without identifying the treatment effect $\EffParam$, which can be challenging under unmeasured confounding.
By contrast, our joint system co-trains $\EffParam$ and $\ParamAlpha$ simultaneously and the baseline function $\EstBase(\cdot;\EffParam,\ParamAlpha)$ adapts to both through the closed-form expression~\eqref{aux-eq:SimpleBaseline}. This has two practical advantages. First, it eliminates the need for a separate no-switching reference group \citep{ying_structural_2023}, as the outcome model is learned from the entire cohort, a feature particularly valuable in observational studies where few patients remain on the initial treatment throughout follow-up. Second, the internal feedback between $\EffParam$ and $(\ParamAlpha,\HazBase)$ ensures that the estimated outcome model is self-consistent with the estimated treatment effect, avoiding the potential incompatibility that arises otherwise.
Together, these properties make the semi-parametric DRIVE estimator well suited to observational treatment-switching studies where a clean no-switching control arm may be limited or difficult to define.

\csubsection{ML-based Estimator}\label{sec-theory-ml}

Define the asymptotic limit of $\cfHaz$, i.e. the cumulative hazard of event time conditionally on baseline covariates within the no-switching control set,
$\HazCtrlTrue(\TimeGrid;\Covs_i) = -\log\left\{
\Prob\left(\TimeSurv_i \ge \TimeGrid \mid \Covs_i, \TreatHist_i = \ctrl{M}\right)\right\}$.
For flexible models, we measure the estimation
error of out-of-fold  machine learning  estimators $\cfPScore$ and $\cfHaz$ over the independent in-fold data by the mean-square error:
\begin{align}
    \notag
    \|\cfPScore - \PStrue\|_\pi = & \sqrt{\Exp_*[\{\cfPScore(\Covs_*) - \PStrue(\Covs_*)\}^2]} \\
    \|\cfHaz - \HazCtrlTrue\|_H = & \sqrt{\Exp_*\left[\sup_{\TimeGrid \in [0,\TimeMax]}\{\cfHaz(\TimeGrid;\Covs_*) - \HazCtrlTrue(\TimeGrid;\Covs_*)\}^2\right]},  \label{def:mse}
\end{align}
where $\Covs_*$ represent an independent measured covariate following the same distribution of $\Covs_i$, and the expectation $\Exp_*$ is taken with respect to $\Covs_*$.

We require the machine learning estimators to be consistent at the following rate.
\begin{assumption}[Regularity for machine learning estimator]\label{ap:ml} \phantom{ }
\begin{enumerate}[label = (\roman*), ref = \ref{aux-ap:ml}(\roman*)]
    \item \label{ap:indepL-ml} \textbf{Covariate invariant bias}. Conditional on $\IV$, the residual survival distortion induced by the unmeasured process is homogeneous across the measured covariates $\Covs$ within no-switching control group.
    \begin{equation*}
        \Covs_i \perp\!\!\!\perp \{\ConfUn_i (t): 0\leq t\leq \TimeMax\} \; |\;  Z_i, \TreatHist_i = \ctrl{M}
    \end{equation*}
    \item \label{ap:productrate}
    \textbf{Estimation of nuisance models}.     The cross-fitted nuisance models converges in the following rate,
    \begin{equation*}
\sup_{j = 1,\dots, K} \sqrt{n} \|\cfPScore - \PStrue\|_\pi\|\cfHaz - \HazCtrlTrue\|_H +  \|\cfPScore - \PStrue\|_\pi + \|\cfHaz - \HazCtrlTrue\|_H = o_p(1).
\end{equation*}
\end{enumerate}
\end{assumption}
For Assumption~\ref{aux-ap:indepL-ml}, we assume that, within levels of $\IV$, the unmeasured process induces a common residual survival-scale distortion among subjects following the control reference path. The measured covariates $\Covs$ may affect the outcome process flexibly through the nuisance hazard, but they do not further modify this residual latent correction. This is needed when we estimate $\Haztrue$ using only no-switching control group.
Assumption~\ref{aux-ap:productrate} imposes the standard requirement that the product of their estimation errors in mean squared error \eqref{aux-def:mse} converge to zero faster than  $n^{-1/2}$. This allows each nuisance component to converge at an arbitrary, possibly slow nonparametric rate, provided that the other converges sufficiently fast, while still ensuring that $\EffEstML$ remains asymptotically unbiased, root-$n$ consistent, and asymptotically normal.
Assumption~\ref{aux-ap:productrate} plays a role analogous to the standard requirement in the semi-parametric literature that each nuisance estimator converge at rate $n^{-1/4}$ or faster \citep{7ee01c69-5c6e-3364-949c-e021536cabe3}, and it is comparable to the rate conditions commonly imposed in double machine learning frameworks \citep{Chernozhukov.Chetverikov.ea2018b, Chen.Huang.ea2021}.

\begin{theorem}\label{thm:ml}
Under Assumptions~\ref{aux-ap:causal}~and~\ref{aux-ap:ml}, the cross-fitted estimator $\EffEstML$ obtained from Algorithm~\ref{aux-alg:one} is consistent and asymptotically normal:
\[
\sqrt{n}\,(\EffEstML - \EffTrue)
\;\;\rightsquigarrow\;\;
N\!\left(0,\, \mathcal{V}_{\mathrm{ML}}\right)
\text{ and } \sqrt{n/\hat{\mathcal{V}}_{\mathrm{ML}}}\,(\EffEstML - \EffTrue)
\;\;\rightsquigarrow\;\;
N\!\left(0,\, 1\right).
\]
\end{theorem}
The proof of Theorem~\ref{aux-thm:ml} adapts the framework originally proposed in \citet{Chernozhukov.Chetverikov.ea2018b} to our model \eqref{aux-eq:SCSM}. Cross-fitting ensures that the nuisance estimates used in each fold are independent of the evaluation sample, so the analysis on the score conditionally on out-of-fold training data takes a simple form of averages of independent and identically distributed variables.
We separated estimation errors of the scores into model error from out-of-fold training and uncertainty from averaging over in-fold data. Neyman orthogonality (Lemma~\ref{aux-lm:orthogonality}) then guarantees that model errors from out-of-fold training enter only through second-order product terms, which vanish faster than $n^{-1/2}$ under Assumption~\ref{aux-ap:productrate}. The estimating equation with estimated nuisances is therefore asymptotically equivalent to the oracle equation evaluated at the true $(\PStrue,\Haztrue)$, to which classical asymptotic analysis applies. A detailed proof is provided in Section \ref{aux-proof:thm-ml} of the Supplementary Materials.

While standard double machine learning primarily corrects regularization bias \citep{Chernozhukov.Chetverikov.ea2018b}, our framework additionally resolves a profound structural bias inherent to survival data, applying ML directly to observational data with unmeasured confounding systematically underestimates the true conditional hazard (see Section~\ref{aux-sec:supp_ml_bias} of Supplementary Materials). Our framework uniquely addresses this by not forcing the ML algorithm to directly learn the true structural hazard $\Haztrue$. Instead, we allow the ML model to flexibly capture a potentially biased proxy $\cfHaz$. The DRIVE then utilizes the IV  to construct the shared empirical profile $\EstBase_r$, which acts as a structural bias-correction term. This mechanism simultaneously profiles out the unmeasured frailty bias and robustifies against ML regularization errors, demonstrating a profound synergy between semiparametric survival theory and modern machine learning.

With potentially non-regular machine learning estimators, we need to always identify the measured confounding effect $\Haztrue$ under Assumption~\ref{aux-ap:indepL-ml} along with the propensity score $\PStrue$ to attain the Neyman orthogonality (see Remark \ref{aux-remark:exp-score}).
The use of effective machine learning estimators, however, may enhance the feasibility of
Assumption~\ref{aux-ap:indepL-ml} by capturing most confounding effects measurable at baseline.

\csection{Numerical Results}\label{sec-sim}
We evaluate DRIVE under two simulation scenarios: one where unmeasured confounding has a mild additive effect on both hazard and switching (Scenario I), and one where switching is driven directly by potential outcomes, making treatment changes prognosis-driven (Scenario II).
For each scenario, we evaluate DRIVE under four configurations: both models correctly specified (Correct), propensity score misspecified (Propensity), survival model misspecified (Survival), and both misspecified (Misspecified).

\csubsection{Data Generation}
For $i=1,\dots,N$, we generate measured covariates $\Covs_i \in \Real^{2}$ and unmeasured confounder $\ConfUn_i \in \Real$ independently from $\mathcal{U}(0,1)$. The binary initial treatment $\IV_i$ is drawn from propensity score model $\Exp(\IV_i \mid \Covs_i) = \PScore(\ParamGamma_0, \Covs_i)$, where different forms of $\PScore(\cdot)$ induce propensity model misspecification. We allow at most one switch per individual at time $W_i$,
\begin{equation*}
    \Treat_i(\TimeGrid) = \IV_i\Indicator(\TimeGrid < W_i) + (1-\IV_i)\Indicator(\TimeGrid \ge W_i).
\end{equation*}

\subsubsection*{Scenario I: Weekly Correlated Switching}
We generate switching times from a semi-parametric additive hazard model.
Given $(\IV_i,\Covs_i,\ConfUn_i)$, the switching time $W_i$ follows an exponential distribution with survival function
\begin{equation}
    \Prob(W_i > \TimeGrid|\IV_i, \Covs_i, \ConfUn_i) =    \exp\{-0.05\TimeGrid +0.25 \IV_i\TimeGrid - 0.5 \Indicator(\IV_i =0)\TimeGrid- \bm\beta_1^\top \Covs_i \TimeGrid - \beta_2 \ConfUn_i\TimeGrid\},\nonumber
\end{equation}
where $(\bm\beta_1^\top, \beta_2)^\top  = (0.125, 0.125, 0.125)$. We generate the survival time $\TimeSurv_i$ from
\begin{equation}
    \label{eq:generationT}
    \Prob(\TimePot[\TreatHist_i] > \TimeGrid|\IV_i, \TreatHist_i, \Covs_i, \ConfUn_i) =  \exp\left\{-0.1\TimeGrid - 0.1\int_0^{\TimeGrid-}\Treat_i(s)ds - \nu (\ParamAlpha, \Covs_i, \ConfUn_i)\TimeGrid\right\},
\end{equation}
where we assume the true treatment effect $\EffTrue = 0.1$. Different forms of $\nu(\cdot)$ induce outcome regression misspecification.

\subsubsection*{Scenario II: Endogenous Switching}
For each patient, we will first generate initial assignment $\IV_i$ and the no-switching potential survival times $\TimePot[\trted{M}]$ and $\TimePot[\ctrl{M}]$. Then, we generate the switching time $W_i$ from
\begin{equation*}
    \Prob(W_i > \TimeGrid|\IV_i, \Covs_i) = \left\{
    \begin{array}{ll}
        \exp\{2\exp(-\TimePot[\trted{M}](-0.1\TimeGrid + 0.5\TimeGrid - \nu (\ParamAlpha, \Covs_i, \ConfUn_i)\TimeGrid)\}, & \IV_i = 1\\
        \exp\{2\exp(-\TimePot[\ctrl{M}](-0.1\TimeGrid - (1 - 0.5)\TimeGrid - \nu (\ParamAlpha, \Covs_i, \ConfUn_i)\TimeGrid)\},& \IV_i = 0.
    \end{array}\right.
\end{equation*}

Patients with larger $\TimePot[\bm{0}_M]$ or $\TimePot[\ctrl{M}]$ tend to switch near the end of follow-up, whereas those with shorter counterfactual survival are more likely to switch earlier. We aim to emulate the clinical decision process, where clinicians tend to maintain a therapy if it indicates good prognosis and switch otherwise.

To generate $\TimePot[\TreatHist_i]$, we impose a rank-preserving assumption, i.e. all potential survival times are at the same quantile of their respective distribution. Under the model \eqref{aux-eq:generationT} for $\TimePot[\TreatHist_i]$ and rank-preserving,
we may obtain $\TimePot[\TreatHist_i]$ from $\TimePot[\ctrl{M}]$ through equation
\[
0.1\TimePot[\TreatHist_i]\int_0^{\TimePot[\TreatHist_i]} \Treat_i(s)\,ds
	\cdot	[0.1 + \nu(\ParamAlpha,\Covs_i,\ConfUn_i)]\{\TimePot[\TreatHist_i] - \TimePot[\ctrl{M}]\} = 0.
\]

\subsubsection*{Propensity and Survival Models}
\begin{equation*}
\PScore(\ParamGamma_0,\Covs_i)=\expit(\ParamGamma_0^\top \Covs_i), \quad \nu(\ParamAlpha_0,\Covs_i,\ConfUn_i)= (\Covs_i^\top,\ConfUn_i)\ParamAlpha_0,
\end{equation*}
with $\ParamGamma_0 = (1, -1)^\top$ and $\ParamAlpha_0 = (0.25, 0.25, 0.25)^\top$. Model misspecification is induced via nonlinear covariate transformations:
\begin{equation*}
\tilde{L}_{i,1}=\Indicator(L_{i,1} \le 0.5) e^{L_{i,1}/2}/4 + \Indicator(L_{i,1}>0.5)\{e^2-e^{L_{i,1}/2}\}, \quad \tilde{L}_{i,2}=\frac{L_{i,2}}{1+\exp(L_{i,1})}+1.
\end{equation*}
Under propensity misspecification, $\PScore(\ParamGamma_0,\Covs_i)=\expit(-4.9 + 0.5\tilde{L}_{i,1})$; under survival misspecification, $\nu(\ParamAlpha_0,\Covs_i,\ConfUn_i)=0.1(|\tilde{L}_{i,1}| + \tilde{L}_{i,2} + \ConfUn_i)$; under both, the transformed covariates are used in both generating models. All simulations use $1{,}000$ replications with $N\in\{1600,3200\}$.

\csubsection{DRIVE and Benchmark Methods}\label{sec-sim-methods}

\textbf{DRIVE semi-parametric ($\hat\EffParam_{\mathrm{joint}}$).} We solve the joint estimating equations~\eqref{aux-eq:jointscore} via logistic regression for $\EstPScore(\Covs_i)$ and a linear additive hazard model for the outcome. The survival model is correctly specified under ``Correct'' and ``Propensity'', and misspecified under ``Survival'' and ``Misspecified'', with the reverse holding for the propensity model.

\textbf{DRIVE machine learning ($\hat\EffParam_{\mathrm{ML}}$).} We apply Algorithm~\ref{aux-alg:one} with $K=10$ folds, estimating the cumulative hazard $\cfHaz$ via random survival forest implemented in the \texttt{randomForestSRC} package \citep{Ishwaran.Kogalur2007} and the propensity score $\cfPScore$ via classification tree using the \texttt{rpart} package \citep{Therneau.Atkinson2023rpart} on training folds, with cross-fitted scores evaluated on held-out folds.

\textbf{Benchmark methods.}
We compare against four benchmarks: a time-varying additive hazard model (TVAH) \citep{LinYing94} that treats observed treatment as exogenous, and three naïve strategies that ignore informative switching—ITT, which models the effect of initial assignment; PP, which restricts to non-switchers; and recensoring, which treats switching time as censoring.
The TVAH estimator is fitted with the Aalen additive hazard model via the \texttt{timereg} package \citep{Martinussen.Scheike2006}.

\csubsection{Simulation results}

In Scenario I (Table \ref{aux-tb:simulation1}), where treatment switching is only weakly correlated with survival, both DRIVE estimators perform well whenever at least one of the nuisance models is correctly specified. They remain generally unbiased, the estimated standard errors are consistent, and the coverage rates stay controlled at the nominal level. Their empirical variability also decreases with larger sample size, reflecting the asymptotic consistency. When both the propensity score and survival models are nonlinear, only the ML-based DRIVE estimator continues to deliver accurate estimation and inference, whereas all competing methods exhibit substantial bias and near-zero coverage. Because switching is only weakly informative in this scenario, the re-censored analysis and the time-varying additive hazard model behave reasonably well under the first three settings. In contrast, the per-protocol approach remains biased, potentially due to selection induced by excluding switchers, while the ITT analysis dilutes the effect toward the null.

In Scenario II (Table \ref{aux-tb:simulation2}), where switching is strongly associated with prognosis, only the two DRIVE estimators remain reliable under the first three settings, maintaining negligible bias, consistent standard-error estimation, and nominal coverage. Their standard deviations again decrease with increasing sample size, supporting the expected convergence behavior. In the fully nonlinear case, the ML-based DRIVE estimator performs best and is the only method that continues to provide valid inference. The strong informativeness of switching leads to severe bias in both the time-varying hazard model and the re-censored approach. The ITT and per-protocol analyses also fail that ITT attenuates the effect, while per-protocol induces selection on outcome risk and thus is biased.

\begin{table}[htbp]
    \centering
    \caption{Simulation results under Scenario I: this scenario corresponds to weakly informative switching. The compared estimators include: PP (per-protocol analysis excluding switchers), ITT (intention-to-treat analysis), TVAH (time-varying additive hazard model), and Recensor (recensored analysis treating switching time as a censoring time). For each method, we report the bias, empirical standard deviation (SD), square root of mean SEs ($\sqrt{\text{MSE}}$), and coverage of the 95\% confidence interval (Coverage).}
    \label{tb:simulation1}
    \resizebox{\textwidth}{!}{\spacingset{1}
        \begin{tabular}{ccccccccc}
            Settings & Sample size & Metrics & $\hat\EffParam_{\mathrm{joint}}$ & $\EffEstML$ & $\hat{\EffParam}_{\text{PP}}$ & $\hat{\EffParam}_{\text{ITT}}$ & $\hat{\EffParam}_{\text{TVAH}}$ & $\hat{\EffParam}_{\text{Recensor}}$\\
\toprule
\multirow{8}{*}{Correct}
 & \multirow{4}{*}{$N = 1,600$} & Bias & -$0.0021$ & -$0.0016$ & $0.3901$ & -$0.0413$ & $0.0007$ & -$0.0018$ \\
 & & SD & $0.0602$ & $0.0629$ & $0.0636$ & $0.0364$ & $0.0423$ & $0.0442$ \\
 & & $\sqrt{\text{MSE}}$ & $0.0592$ & $0.0607$ & $0.0600$ & $0.0357$ & $0.0412$ & $0.0430$ \\
 & & Coverage & 0.944 & 0.942 & 0.000 & 0.775 & 0.941 & 0.948 \\
 \cline{2-9}
 & \multirow{4}{*}{$N = 3,200$} & Bias & $0.0004$ & $0.0003$ & $0.3884$ & -$0.0401$ & -$0.0004$ & -$0.0017$ \\
 & & SD & $0.0418$ & $0.0426$ & $0.0440$ & $0.0250$ & $0.0300$ & $0.0310$ \\
 & & $\sqrt{\text{MSE}}$& $0.0419$ & $0.0425$ & $0.0423$ & $0.0252$ & $0.0291$ & $0.0304$ \\
 & & Coverage & 0.954 & 0.950 & 0.000 & 0.651 & 0.944 & 0.951 \\
\midrule
\multirow{8}{*}{Propensity}
 & \multirow{4}{*}{$N = 1,600$} & Bias & $0.0005$ & $0.0019$ & $0.3995$ & -$0.0401$ & -$0.0004$ & -$0.0008$ \\
 & & SD & $0.0727$ & $0.0788$ & $0.0708$ & $0.0433$ & $0.0460$ & $0.0502$ \\
 & & $\sqrt{\text{MSE}}$ & $0.0691$ & $0.0757$ & $0.0674$ & $0.0415$ & $0.0452$ & $0.0496$ \\
 & & Coverage & 0.944 & 0.942 & 0.000 & 0.830 & 0.953 & 0.950 \\
 \cline{2-9}
 & \multirow{4}{*}{$N = 3,200$} & Bias & -$0.0005$ & -$0.0016$ & $0.3977$ & -$0.0406$ & -$0.0010$ & -$0.0020$ \\
 & & SD & $0.0496$ & $0.0540$ & $0.0475$ & $0.0297$ & $0.0315$ & $0.0349$ \\
 & & $\sqrt{\text{MSE}}$ & $0.0484$ & $0.0528$ & $0.0474$ & $0.0291$ & $0.0318$ & $0.0348$ \\
 & & Coverage & 0.953 & 0.946 & 0.000 & 0.692 & 0.958 & 0.946 \\
 \midrule
\multirow{8}{*}{Survival}
 & \multirow{4}{*}{$N = 1,600$} & Bias & $0.0009$ & $0.0028$ & $0.4391$ & -$0.0417$ & $0.0153$ & $0.0120$ \\
 & & SD & $0.0697$ & $0.0589$ & $0.0763$ & $0.0404$ & $0.0498$ & $0.0517$ \\
 & & $\sqrt{\text{MSE}}$ & $0.0724$ & $0.0625$ & $0.0724$ & $0.0373$ & $0.0483$ & $0.0500$ \\
 & & Coverage & 0.959 & 0.960 & 0.000 & 0.789 & 0.936 & 0.932 \\
 \cline{2-9}
 & \multirow{4}{*}{$N = 3,200$} & Bias & -$0.0010$ & $0.0022$ & $0.4362$ & -$0.0429$ & $0.0142$ & $0.0104$ \\
 & & SD & $0.0520$ & $0.0431$ & $0.0556$ & $0.0301$ & $0.0361$ & $0.0382$ \\
 & & $\sqrt{\text{MSE}}$ & $0.0512$ & $0.0438$ & $0.0509$ & $0.0263$ & $0.0341$ & $0.0352$ \\
 & & Coverage & 0.951 & 0.952 & 0.000 & 0.604 & 0.918 & 0.920 \\
\midrule
\multirow{8}{*}{Misspecified}
 & \multirow{4}{*}{$N = 1,600$} & Bias & $0.3986$ & $0.0008$ & $0.9410$ & $0.2768$ & $0.3396$ & $0.4170$ \\
 & & SD & $0.1587$ & $0.0749$ & $0.1051$ & $0.0529$ & $0.0672$ & $0.0679$ \\
 & & $\sqrt{\text{MSE}}$ & $0.0876$ & $0.0780$ & $0.0899$ & $0.0459$ & $0.0631$ & $0.0631$ \\
 & & Coverage & 0.003 & 0.962 & 0.000 & 0.000 & 0.001 & 0.000 \\
 & \multirow{4}{*}{$N = 3,200$} & Bias & $0.3917$ & -$0.0002$ & $0.9444$ & $0.2762$ & $0.3391$ & $0.4159$ \\
 & & SD & $0.0583$ & $0.0566$ & $0.0781$ & $0.0419$ & $0.0508$ & $0.0519$ \\
 & & $\sqrt{\text{MSE}}$ & $0.0572$ & $0.0547$ & $0.0635$ & $0.0324$ & $0.0445$ & $0.0445$ \\
 & & Coverage & 0.000 & 0.948 & 0.000 & 0.000 & 0.000 & 0.000 \\
\bottomrule
        \end{tabular}}
\end{table}

\csection{Real Data Application}\label{sec-data}
We applied the proposed DRIVE and all benchmark methods described in Section \ref{aux-sec-sim-methods} to the study of standard efficacy DMTs versus high efficacy DMTs as second line treatment for MS using  Comprehensive Longitudinal Investigation of Multiple Sclerosis at Brigham and Women's Hospital (CLIMB). Starting from 2000, the CLIMB is an ongoing prospective study of Multiple Sclerosis (MS), enrolling participants with neurologist-diagnosed MS, $>$18 years old. It collects patient-level baseline features, MS relapses and DMTs during the follow-up. During the entire study period, participants will undergo serial mono-therapies of DMTs, and possibly experience treatment switching because of treatment failure due to relapse. While most patients started standard efficacy DMTs as their first line treatment due to affordability, accessibility and mild adverse effect profile, there is a knowledge gap in how the choice of subsequent treatment after switching affect future disease activity. Our analysis therefore compares the overall effectiveness of high-efficacy versus standard-efficacy DMT strategies while incorporating treatment switching, which may provide supportive evidence for subsequent treatment decisions. Time is measured in units of year.

We included participants who initiated their second DMTs between January 1, 2006 and December 31, 2016.
We excluded patients who initiated treatment before 2006 due to lack of overlapping with limited approved high-efficacy DMTs at the time. We also linked the EHR data for each participant from the Mass General Brigham HealthCare System to derive demographic variables (ie, age, sex, race, and ethnicity), clinical variables (ie, disease, follow-up, and prior DMT use duration and number of relapses in the prior 1 and 2 years), and other expert-defined EHR features as confounders \citep{Hou.Kim.ea2021}. The study cohort contained 795 patients, among which 208 patients (26.1\%) experienced treatment switching and 324 patients (40.7\%) experienced relapsing during the study period. We regarded the group of DMT at baseline as the IV with 95 patients (11.9\%) initialized with high efficacy, of which $54.7\%$ experienced treatment switching, and 700 patients (88.1\%) initialized with standard efficacy, of which $22.3\%$ experienced treatment switching. There are 544 patients (68.4\%) in the no-switching control set who started standard efficacy DMT at baseline and never switched to a high efficacy DMT throughout the follow-up.
We applied the same six estimators described in Section~\ref{aux-sec-sim-methods} to estimate the causal effect of high-efficacy DMT on time to next relapse.



Table \ref{aux-tb:Real} summarizes the results of the real-data analysis, including point estimates, standard errors and p-values from DRIVE and benchmark estimators. Several patterns emerge that provide insight into both the treatment effect and the behavior of commonly used analytic strategies.
With the exception of the per-protocol analysis, all methods yield negative point estimates, suggesting a potential reduction in relapse risk associated with a high-efficacy DMT. The per-protocol estimator is the only one that produces a positive estimate, likely reflecting selection bias introduced by restricting analysis to patients who never switched treatment.
The two proposed DRIVE estimators, $\EffEstJoint$ and $\EffEstML$, produce similar results, indicating that model misspecification, if present, is not severe. Both estimators show a negative effect estimate but with P value $> 0.05$.
The estimated treatment effect of the ITT analysis was closer to zero than those from DRIVE potentially due to the dilution of treatment effect.
The time-varying additive hazard model and the recensored analysis suggested a significant lower risks for high efficacy DMTs with a comparatively smaller standard error. These methods do not account for potential unmeasured confounding and therefore may underestimate uncertainty. However, their point estimates remain close to those of our proposed methods.
The evidence from all these methods suggests that a high-efficacy DMT may reduce relapse risk, though a larger study would be needed to achieve sufficient power. The variability of results across methods also highlights the importance of addressing treatment switching and unmeasured confounding when analyzing real-world longitudinal treatment data.

\begin{table}[ht]
\centering
\caption{Summary of real data analysis. Methods are applied to the CLIMB dataset, which tracks MS patients receiving high- or standard-efficacy DMTs between 2006 and 2016. The table presents the estimated effects of high-efficacy DMT on relapse rates in 796 participants, including 208 (26.1\%) who experienced treatment switching and 324 (40.7\%) who experienced relapses. Our methods, $\EffEstJoint$ and $\EffEstML$ (using random forest and 10-fold cross-fitting) use joint estimation and cross-fitting machine learning integration.}
\label{tb:Real}
{\spacingset{1}
\begin{tabular}{lccc}
  \toprule
  \textbf{Method}            & \textbf{Estimate ($\hat{\EffParam}$)} & \textbf{SE} & \textbf{Pval} \\
  \midrule
  $\EffEstJoint$           & -0.0274  & 0.0192 & 0.1546 \\
  $\EffEstML$           & -0.0324  & 0.0208 & 0.1195 \\
  ITT (intention-to-treat)                                     & -0.0186  & 0.0127 & 0.1444 \\
  Recensoring        & -0.0341  & 0.0151 & 0.0240$^{*}$ \\
  Per-Protocol         & 0.0085   & 0.0221 & 0.7021 \\
  Time-Varing Additive Hazard         & -0.0261  & 0.0123 & 0.0338$^{*}$ \\
  \midrule
  \multicolumn{4}{r}{$^{*} p < 0.05$, $^{**} p < 0.01$, $^{***} p < 0.001$} \\
  \bottomrule
  \end{tabular}}
\end{table}

\csection{Discussion}\label{sec-conc}

We proposed DRIVE, a doubly robust IV method for estimating treatment effects on time-to-event outcomes in observational studies with potential informative treatment switching. By framing unmeasured reasons for switching as time-varying confounding, we recognized the initial treatment as an IV and identify the causal effect under a structural cumulative survival model. The resulting estimator is doubly robust under semi-parametric models: it remains consistent and asymptotically normal when either the propensity score for the initial treatment or the measured confounding effect is correctly specified. The semi-parametric version co-trains the treatment effect and the outcome regression via joint estimating equations, avoiding the need for a no-switching reference group. The machine-learning version uses data-driven correction of baseline function and cross-fitting to allow flexible nuisance estimation while preserving root-$n$ consistency and asymptotic normality under a product-rate condition.  Simulations showed that both DRIVE estimators perform well under a range of data-generating processes where common alternatives (ITT, per-protocol, recensored, and time-varying additive hazard analyses) can be biased. The application to DMTs for multiple sclerosis illustrated the practical use of the method when switching is common and a per-protocol population free of  selection bias is hard to define.

A key advantage of EHR-based studies, compared with randomized trials, is the availability of rich longitudinal clinical information collected in routine care. The machine-learning version of DRIVE leverages this advantage by flexibly incorporating baseline and historical information beyond a small set of prespecified covariates. This flexibility may become increasingly valuable as national databases are optimized and linked with information from general practitioners and specialized clinical sources.
At the same time, the flexibility of the machine-learning formulation comes with additional theoretical requirements as Assumption~\ref{aux-ap:ml}. In settings where these requirements are considered restrictive, more structured nonparametric approaches, such as spline- or sieve-based methods, could alternatively be integrated into the semiparametric version of DRIVE. These approaches may yield explicit convergence rates while preserving double robustness, but they are restricted by smoothness or complexity assumptions, which is less aligned with the goal of flexibly leveraging rich EHR information. We view these directions as complementary to the machine-learning DRIVE framework developed here.

In our setting, the baseline (initial) treatment is a natural IV because it is the main source of variation in treatment trajectories, and it can be taken as good-as-random given baseline covariates. A time-varying IV would be needed to increase statistical power when exogenous variation in treatment arises during follow-up rather than only at baseline.
For example, when policy or eligibility changes at specific calendar times create new encouragement to switch, or when treatment is revisited at repeated decision points with quasi-random variation at each time. Related methods already exist for such settings. \citet{Michael.Cui.ea2024} develop IV estimation of marginal structural mean models for time-varying treatment, using time-varying IVs to identify effects when sequential randomization fails due to unmeasured confounding, although it heavily relies on correct specification of the propensity score, which can be difficult to achieve as the complexity of the data and decision process grows.
Extending DRIVE to settings with time-varying IVs could therefore help. The machine-learning version would offer additional flexibility for modeling complex propensity and outcome structures in such settings.

The SCSM model used targets the average treatment effect in the study population, but this effect may obscure clinically important heterogeneity and provide incomplete guidance. Evaluating effect modification therefore helps translate the causal analysis from a population-level comparison to a clinically actionable assessment of which patients are most likely to benefit from each treatment strategy.
Extending DRIVE to estimate effect modification by baseline covariates would require modeling $\EffParam(\Covs)$ or similar summaries and adapting the score and variance estimation. \citet{Almirall_EffectModification_2010} studied effect modification in structural nested mean models for time-varying treatment under a prespecified linear modification structure. A similar parametric extension could be incorporated into DRIVE while preserving double robustness under analogous nuisance-model conditions. In addition, the machine-learning version of DRIVE could provide a more flexible framework for estimating nonlinear or high-dimensional effect modification patterns.

When the primary endpoint is rare or requires long follow-up, surrogate endpoints are often used. Formal criteria for valid surrogacy and methods that use surrogates to infer effects on the primary outcome have been studied extensively \citep{Frangakis.Rubin2002}. DRIVE is naturally extendable to this setting. The outcome model already conditions on measured covariates to capture the hazard of the potential outcome, so the surrogate can be incorporated as an additional covariate in that model. The same IV continues to identify the treatment effect on the primary time-to-event endpoint and double robustness is preserved. \citet{hou_efficient_2025} derive efficient semi-supervised ATE estimation with partially annotated treatment and response and noisy surrogates in EHR, and \citet{zhang_double_2023} establish double robust semi-supervised inference for the mean under MAR labeling with decaying overlap; both focus on a single treatment decision (point treatment) and share with DRIVE the goal of leveraging auxiliary or surrogate information while retaining robustness. Extending DRIVE to incorporate surrogates would therefore add the setting of time-varying treatment and informative switching, where a single treatment point cannot capture the exposure trajectory.

  \bibliography{bibliography_cleaned}

\end{document}